\documentclass[12pt]{book}

\usepackage[dvips]{graphicx,color}
\usepackage{makeidx,tsukuba}
\newcommand{\lsim}{\mathrel{\mathop{\kern 0pt \rlap
  {\raise.2ex\hbox{$<$}}}
  \lower.9ex\hbox{\kern-.190em $\sim$}}}

\makeauthorindex
\makeindex

\begin{document}

\BookTitle{\itshape The 28th International Cosmic Ray Conference}
\CopyRight{\copyright 2003 by Universal Academy Press, Inc.}
\pagenumbering{arabic}

\chapter{
Stable and Radioactive Nuclei in a Diffusion Model }

\author{%
%
%
Fiorenza Donato,$^1$ David Maurin,$^2$ and Richard Taillet$^3$ \\
{\it (1) Dipartimento di Fisica Teorica, University of Torino, 
10125 Torino, Italy\\
(2) Service d'Astrophysique, SAp CEA-Saclay, 91191 Gif-sur-Yvette, France\\
(3) LAPTH and Universit\'e de Savoie, 74941 Annecy-le-Vieux, France \\}
}

\section*{Abstract}
We present the results on the source spectrum function for 
primary nuclei in galactic cosmic rays, where two distinct energy dependences 
are used for the source spectra. We discuss  the evolution of the goodness of 
fit to B/C data with the propagation  parameters and also show that the 
results are not much affected by a different choice for the diffusion 
scheme. 
We apply the constraints on the diffusion scheme as derived from stable nuclei
to calculate the propagation of beta-radioactive isotopes. The diffusion 
model is refined to properly take into account the effect of a local bubble
in the interstellar medium. 
Our calculations are compared to existing data, which  prefer 
the local bubble description instead of the homogeneus scheme. 

\section{Introduction}
Galactic cosmic rays detected with energies from 100 MeV/nuc to 100 GeV/nuc
were most probably
produced by the acceleration of a low energy galactic population of
nuclei and diffused  by  the turbulent magnetic field.
The acceleration and diffusion processes have a magnetic
origin, so that they should depend on rigidity.
The rigidity function for the diffusion coefficient is given by
quasi-linear theory as
\begin{equation}
    K({\cal R}) = K_0 \beta  \left( \frac{\cal R}{\mbox{1 GV}} \right)^\delta
\end{equation}
where the parameters $K_0$ and $\delta$ should ideally be given by the
small-scale structure of the magnetic field responsible for the
diffusion. 
The spectrum just after acceleration and for a species $j$, $Q^j({\cal R})$, 
depends on the details of the  acceleration process, which are not clearly 
understood. However, a power--law seems to be preferred [1] and we 
assumed the following two possible forms:
\begin{equation}
          Q^j({\cal R}) = \frac{q^j_0}{\beta} \left(\frac{{\cal R}}{1
          \mbox{GV}}\right)^{-\alpha}
          \label{pure_power_law}
\end{equation}
\begin{equation}
          Q^j({\cal R}) = q^j_0 \left(\frac{{\cal R}}{1
\mbox{GV}}\right)^{-\alpha}\;.
\label{modif_power_law}
\end{equation}
where the value of $\alpha$ is still debated (grossly from about 1.5 to
2.5).
\\
We propagate accelerated particles in a two-zones diffusion model, which 
has been described at length in [1, 2]. Briefly, our model takes into account 
spatial diffusion and convection wind ($V_c$), both acting in the thin matter 
disc and in the diffusive halo (this last has unknown size $L$ kpc). 
Spallations over the interstellar matter, reacceleration ($V_a$)
 and standard  energy losses are set in the thin disc. 
The complete diffusion equation and its solutions may be found in [1] 
(see Eq. 4).

\section{Analysis of stable nuclei}
\begin{figure}[t]
  \begin{center}
\includegraphics[height=16.5pc]{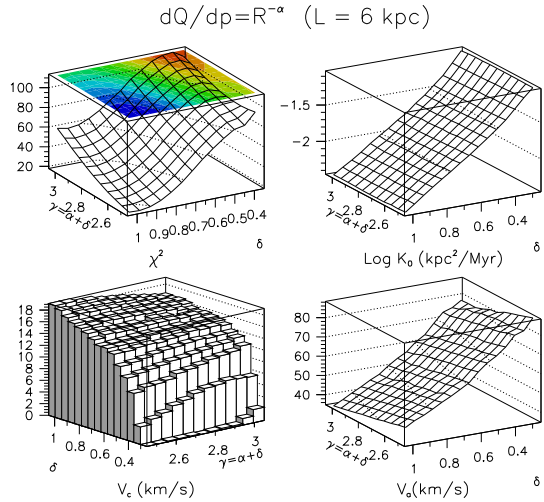}
  \end{center}
  \vspace{-0.5pc}
  \caption{From top to bottom: for each best $\chi^2$ in the plane 
$\delta-\gamma$ ($L=6$ kpc), the corresponding values of $\log(K_0)$, $V_c$ 
and $V_a$ are plotted.}
\label{f3_stables}
\end{figure}

We tested our diffusion model by using data on the B/C ratio.
The analysis was performed with six free parameters: 
$K_0, \delta, V_C, V_a, L, \alpha$ and with the implementation 
of both the source spectra alternatively.
We produced all the chain of primary and secondary nuclei by strating from 
Sulfur. Our aim is to compare predictions for the B/C spectrum with existing 
measurements, in particular with the 26 data points from {\sc heao3} [3].
\\
In Fig.~\ref{f3_stables} we show the preferred values of the three
diffusion parameters $K_0$, $V_c$ and $V_a$, for each best $\chi^2$ in the
$\delta-\gamma$ plane. Here we defined $\gamma = \delta + \alpha$. 
$L$ has been fixed to 6 kpc but we checked that the behavior does not 
particulary depend on $L$. The source spectrum is the pure power law one of 
Eq. (\ref{pure_power_law}).
The two upper panels show that the evolution of $\alpha$ does not affect $K_0$.
On the other hand, we clearly see the (anti)correlation
between $K_0$ and $\delta$ entering the diffusion
coefficient formula, because they need to give about the same
normalization at high energy ($K_0\times E_{\rm thresh}^\delta\approx cte$).
The lower left panel shows the values for the convective velocity. Only very
few configurations include $V_c=0$~km~s$^{-1}$, always when $\delta=0.3$ (for
both types of source spectra). Increasing $\gamma$ and $\delta$ at
the same time makes $V_c$ change its trend. 
For small diffusion slope, convection is unfavored,
as found in [4]. Finally,
the Alfv\'en  velocity (lower right panel) $V_a$ doubles from 
$\delta = 1.0$ to 0.3, whereas it
is almost unchanged by a variation in the parameter $\gamma$.
We found in [1] that the three parameters $K_0$, $V_c$ and $V_a$
behave very similarly with respect to a change in the source spectrum
from ``pure power law" to ``modified". In fact, its influence
on the primary and secondary fluxes can be factored out if
energy changes are discarded.
The currently available data on B/C do not allow to discriminate
clearly between these two shapes for the acceleration spectrum.

\section{Radioactive isotopes in the Local Bubble}
\begin{figure}[h]
  \begin{center}
\includegraphics[height=14.5pc]{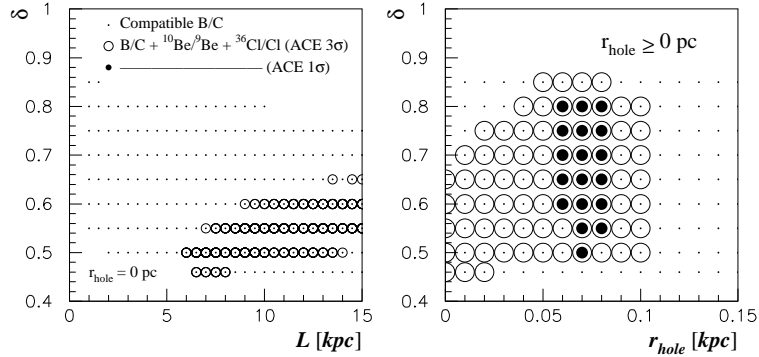}
  \end{center}
  \vspace{-0.9pc}
  \caption{Representation of the models compatible with B/C plus both
$^{10}$Be/$^9$Be and $^{36}$Cl/Cl {\sc ace} 3-$\sigma$ (open
circles) and 1-$\sigma$ (filled circles) [7]. Left panel displays homogeneous 
models ($r_{\rm hole}=0$) in the plane $L-\delta$.
Right panel displays inhomogeneous models ($r_{\rm hole}\geq  0$) in the
plane $r_{\rm hole}-\delta$.}
\label{fig:be_et_cl}
\end{figure}

We studied the compatibility of our diffusion model  with current data 
on $\beta$-radioactive isotopes. These species diffuse on a typical distance 
$l_{\rm rad}\equiv \sqrt{K\gamma \tau_0}$ before decaying.
In this expression, not only the diffusion coefficient $K$, but also
the lifetime $\gamma \tau_0$, depends on energy, due to the relativistic
time stretch.
In Table \ref{tab:rad} we give the values of some $l_{\rm rad}$
for  $^{10}$Be, $^{26}$Al and  $^{36}$Cl. 
\begin{table}[hbt!]
          \begin{center}
	\begin{tabular}{|c|c|cc|}   \hline
	    & $\tau_0$ (Myr) & 1 GeV/nuc  & 10 GeV/nuc \\ \hline
	    $^{10}$Be & 2.17  & 220 pc &  950 pc\\
	    $^{26}$Al & 1.31  & 110 pc & 470 pc\\
	    $^{36}$Cl & 0.443 &  56 pc &  250 pc \\
	    \hline
	\end{tabular}
\caption{Rest frame lifetimes and corresponding values of $l_{rad}$ for
some $\beta$ radioactive  nuclei at two different energies 
( $K_0$ = 0.033 kpc$^2$  Myr$^{-1}$ and $\delta = 0.6$).}
          \label{tab:rad}
          \end{center}
\end{table}
These species are therefore very sensitive to the characteristics of the local 
interstellar medium ({\sc lism}). The Solar System is 
embedded in an underdense region, usually called the Local Bubble (see [5] 
for references). The bubble leads substantially to a decrease 
in the spallation source term of the radioactive species.
We modelled the  bubble as a hole in the thin disc approximation and 
the radius of this hole is considered as an unknown parameter in the analysis.
In [5] we provided the theoretical tools to treat the presence 
of the hole in the galactic disc in our two-zone diffusion model. 
The major result we found is that at the center of
the bubble, the radioactive fluxes are decreased as
\begin{displaymath}
       \frac{N^{r_{\rm hole}}}{N^{r_{\rm hole}=0}}\propto \exp(-r_{\rm
hole}/l_{\rm rad})\;.
\end{displaymath}
\\
Using the diffusion parameters allowed by the B/C data [2] to
compute the $^{10}$Be/$^9$Be and $^{36}$Cl/Cl ratios,
we find that each of the radioactive nuclei points towards
a bubble of radius $\lsim 100$ pc, in relatively good agreement
with direct observations.
If these nuclei are considered simultaneously, only models with a bubble
radius $r_{\rm hole} \sim 60 - 100$ pc are consistent with the data and 
the case for $r_{\rm hole}=0$ pc is disfavored.
This is shown in Fig.~\ref{fig:be_et_cl}, which is a projection of
the parameter subspace allowed by B/C, $^{10}$Be/$^9$Be and $^{36}$Cl/Cl
on the $L-\delta$ plane (left panel, no hole) or $r_{\rm hole}-\delta$
plane (right panel, hole $r_{\rm hole}$).
When the ratio $^{26}$Al/$^{27}$Al is added to the analysis, the results
become less clear [5], and it is suspected that the data 
(nuclear or astrophysical) on which they rely should not be trusted.


\section{References}

\vspace{\baselineskip}

\re
1.\ Maurin D., Taillet R., Donato F. 2002, A\&A 394, 1039
\re
2.\ Maurin D., Donato F., Taillet R., Salati P. 2001, ApJ 555, 585
\re
3.\ Engelmann J.J. et al. 1990, A\&A, 233, 96
\re
4.\ Strong A.W. and Moskalenko I.V. 1998, ApJ, 509, 212
\re
5.\ Donato F., Maurin D., Taillet R. 2002, A\&A 381, 539
\re
6.\ Donato F. et al. 2001, ApJ 563, 172
\re
7.\ Binns W.R. et al., 1999 , ICRC 26 Salt Lake City, OG-1.1.06
\endofpaper
\end{document}